\documentclass[twocolumn,showpacs,preprintnumbers,amsmath,amssymb]{revtex4}

\usepackage{graphicx}
\usepackage{dcolumn}
\usepackage{bm}

\usepackage{amsfonts}
\usepackage{amsthm}
\usepackage{amssymb}
\usepackage[mathscr]{euscript}
\usepackage{t1enc}
\usepackage{amsmath}
\usepackage{amscd}
\usepackage{psfrag}
\usepackage{graphics}

\begin{document}


\title{Does a functional integral really need a Lagrangian?}

\author{Denis Kochan}
\email{kochan@fmph.uniba.sk}
\affiliation{%
Department~of~Theoretical Physics, \\
FMFI, Comenius University, \\
842 48 Bratislava, Slovakia}

\date{\today}

\begin{abstract}
Path integral formulation of quantum mechanics (and also other equivalent formulations) depends on
a Lagrangian and/or Hamiltonian function that is chosen to describe the underlying classical system.
The arbitrariness presented in this choice leads to a phenomenon called \emph{Quantization ambiguity}. For example
both $L_1=\dot{q}^2$ and $L_2=e^{\dot{q}}$ are suitable Lagrangians on a classical level ($\delta L_1=\delta L_2$),
but quantum mechanically they are diverse.

This paper presents a simple rearrangement of the path integral to a surface functional integral. It is shown
that the surface functional integral formulation gives transition probability amplitude which is free of any
Lagrangian/Hamiltonian and requires just the underlying classical equations of motion. A simple example examining
the functionality of the proposed method is considered.\\

\centerline{Dedicated to my friend and colleague Pavel B\'{o}na.}
\end{abstract}


\maketitle

\section{A standard path integral lore}

According to Feynman \cite{feynman1}, the probability amplitude of the transition of a system from the space-time
configuration $(q_0,t_0)$ to $(q_1,t_1)$ is given as follows:
\begin{equation}\label{FeynmanI}
\mathbf{A}(q_1,t_1\,|\,q_0,t_0)\propto
\int[\mathscr{D}\tilde{\gamma}]\exp{\Bigl\{\frac{i}{\hslash}\int\limits_{\tilde{\gamma}}\,p_adq^a-Hdt\Bigr\}}\,.
\end{equation}
Here the path-summation is taken over all trajectories $\tilde{\gamma}(t)=(\tilde{q}(t),\tilde{p}(t),t)$ in the
extended phase space which are constrained as follows:
$$
\begin{aligned}
\tilde{\gamma}(t_0)&=\bigl(\tilde{q}(t_0)=q_0,\tilde{p}(t_0)\mbox{\,-\,arbitrary},t_0\bigr)\,,\\
\tilde{\gamma}(t_1)&=\bigl(\tilde{q}(t_1)=q_1,\tilde{p}(t_1)\mbox{\,-\,arbitrary},t_1\bigr)\,.
\end{aligned}
$$

\noindent To obtain a proper normalization of the Feynman propagator, one requires:
$$
\begin{aligned}
\delta(\tilde{q}_0-q_0)&=\int\limits_{{\mathbf{R}^n[q_1]}}dq_1\,\boldsymbol{\mathbf{A}^*}(q_1,t_1\,|\,\tilde{q}_0,t_0)\,\mathbf{A}(q_1,t_1\,|\,q_0,t_0)\,,\\
\delta(q_1-q_0)&=\lim\limits_{t_1\rightarrow t_0}\mathbf{A}(q_1,t_1\,|\,q_0,t_0)\,.
\end{aligned}
$$
The first equation asks for the conservation of the total probability and the second expresses the obvious fact that no
evolution takes place whenever $ t_1$ approaches $t_0$.

It is a miraculous consequence (not a requirement!) of the propagator definition (\ref{FeynmanI}) that it satisfies an
evolutionary chain rule (Chapman-Kolmogorov equation)
$$
\mathbf{A}(q_1,t_1\,|\,q_0,t_0)=\int\limits_{{\mathbf{R}^n[q]}} dq\ \mathbf{A}(q_1,t_1\,|\,q,t)\,\mathbf{A}(q,t\,|\,q_0,t_0)\,,
$$
whose infinitesimal version is the celebrated Schr\"{o}dinger equation.

\noindent It is a curious fact that Formula (\ref{FeynmanI}) was not originally discovered by Feynman.
In his pioneering paper \cite{feynman2} he arrives at a functional integral in the configuration space only
\begin{equation}\label{FeynmanII}
\mathbf{A}(q_1,t_1\,|\,q_0,t_0)\propto
\int[\mathscr{D}q]\exp{\Bigl\{\frac{i}{\hslash}\int\limits_{q(t)}L(q,\dot{q},t)dt\Bigr\}}\,.
\end{equation}
Later, however, it was shown that this formula represents a very special case of the most general prescription
(\ref{FeynmanI}). Formula (\ref{FeynmanI}) is at the heart of our further discussion.

\section{DeHamiltonianization}

A step beyond involves eliminating the Hamiltonian function $H$ from Formula (\ref{FeynmanI}).
The price to be paid for this will be to replace the path summation therein by a surface functional integration.

\noindent Our aim is the transition probability amplitude between $(q_0,t_0)$ and $(q_1,t_1)$. Let us suppose
that there exists a unique classical trajectory in the extended phase space $\gamma_{cl}(t)=(q_{cl}(t),p_{cl}(t),t)$
which connects these points (locally this assumption is always satisfied).
Then for any curve $\tilde{\gamma}(t)=(\tilde{q}(t),\tilde{p}(t),t)$ which enters the path integration in
(\ref{FeynmanI}), we can assign two auxiliary curves which we call $\lambda_0(s)$ and $\lambda_1(s)$.
They are parameterized by $s\in[0,1]$ and specified as follows:
$$
\begin{aligned}
\lambda_0(s)&=(q_0,\pi_0(s),t_0)\, \mbox{where}\, \pi_0(0)=p_{cl}(t_0),\,\pi_0(1)=\tilde{p}(t_0),\\
\lambda_1(s)&=(q_1,\pi_1(s),t_1)\, \mbox{where}\, \pi_1(0)=p_{cl}(t_1),\,\pi_1(1)=\tilde{p}(t_1).\\
\end{aligned}
$$
Let us emphasize that neither $\lambda_0(s)$ nor $\lambda_1(s)$ varies with respect to the $q$ and $t$ coordinates in
the extended phase space. They are allowed to evolve only with respect to the momentum variables. There are, of course,
infinitely many of such curves, but as we will see nothing in the theory will be dependent on a particular
choice of $\lambda_0(s)$ and $\lambda_1(s)$.

Using these curves one can write:
\begin{equation}\label{loop}
\int\limits_{\tilde{\gamma}} p_adq^a-Hdt=\int\limits_{\gamma_{cl}} p_adq^a-Hdt + \oint\limits_{\partial\Sigma} p_adq^a-Hdt\,,
\end{equation}
where $\partial\Sigma=\tilde{\gamma}-\lambda_1-\gamma_{cl}+\lambda_0$ is a contour spanned by four
curves $\tilde{\gamma}(t)$, $\gamma_{cl}(t)$, $\lambda_0(s)$, $\lambda_1(s)$ counting their orientations.

\noindent The first integral on the right is the classical action $\mathbf{S}_{cl}(q_1,t_1\,|\,q_0,t_0)$. While the contour
integral in (\ref{loop}) can be rearranged to represent a surface integral:
\begin{equation}
\begin{aligned}\label{formula 4}
\oint\limits_{\partial\Sigma} & p_adq^a - Hdt=\\
& \int\limits_{\Sigma} dp_a\wedge\Bigl(dq^a-\frac{\partial H}{\partial p_a}dt\Bigr)-\frac{\partial H}{\partial q^a}dq^a\wedge dt=:\int\limits_{\Sigma}\Omega\,.
\end{aligned}
\end{equation}
Surface $\Sigma$ spanning the contour $\partial\Sigma$ is understood here as a map from the parametric
space $(t,s)\in[t_0,t_1]\times[0,1]$ to the extended phase space, i.e.
$$
\Sigma: (t,s)\mapsto \bigl(q^a(t,s),p_a(t,s),t(t,s)=t\bigr)\,.
$$
Partial derivatives of the initial Hamiltonian function can be substituted using the velocity-momentum relations
and classical equations of motion:
$$
\frac{\partial H}{\partial p_a}=T^{ab}p_b\bigl(=\dot{q}^a\bigr)\ \ \ \mbox{and}\ \ \ \frac{\partial H}{\partial q^a}=-F_a\bigl(=-\dot{p}_a\bigr)\,.
$$
Here we consider the physically mostly relevant situation only. In this case the velocities and momenta become related linearly
by the metric tensor $T_{ab}(q)$ (and its inverse) defined by the kinetic energy
$T=\tfrac{1}{2}T_{ab}\dot{q}^a\dot{q}^b=\tfrac{1}{2}T^{ab}p_ap_b$ of the system, then
\begin{equation}\label{Omega}
\Omega=dp_a\wedge dq^a-\Bigl(T^{ab}p_adp_b -F_adq^a\Bigr)\wedge dt\,.
\end{equation}
This object represents a canonical two-form in the extended phase space. It is a straightforward generalization
of the standard closed two-form $d\theta=dp\wedge dq-dH\wedge dt$ to the case when the forces are not potential-generated.

It is clear that for a given pair of trajectories $(\tilde{\gamma},\gamma_{cl})$ there exists infinitely many $\Sigma$
surfaces. They form a set which we call $\mathscr{U}_{\tilde{\gamma}}$. Since the surface integral $\int_\Sigma\Omega$
is only boundary dependent and Formulas (\ref{loop}) and (\ref{formula 4}) are satisfied, we can write:
$$
\exp{\Bigl\{\frac{i}{\hslash}\int\limits_{\tilde{\gamma}}p_adq^a-Hdt\Bigr\}}
=\frac{\mbox{\large{e}}^{\frac{i}{\hslash}\mathbf{S}_{cl}}}{\mathrm{\infty}_{\tilde{\gamma}}}
\int\limits_{\mathscr{U}_{\tilde{\gamma}}}[\mathscr{D}\Sigma] \exp{\Bigl\{\frac{i}{\hslash}\int\limits_{\Sigma}\Omega\Bigr\}}.
$$
Here $\mathrm{\infty}_{\tilde{\gamma}}$ stands for the number of elements pertaining to the corresponding stringy set
$\mathscr{U}_{\tilde{\gamma}}$. Assuming no topological obstructions from the side of the extended phase space,
$\mathrm{\infty}_{\tilde{\gamma}}$ becomes an infinite constant independent of $\tilde{\gamma}$.
Taking all of this into account we can rewrite (\ref{FeynmanI}) as follows:
\begin{equation}{\label{FeynmanIII}}
\mathbf{A}(q_1,t_1\,|\,q_0,t_0)\propto
\mathrm{e}^{\frac{i}{\hslash}\mathbf{S}_{cl}}\int\limits_{\mathscr{U}}[\mathscr{D}\Sigma]\exp{\Bigl\{\frac{i}{\hslash}\int\limits_{\Sigma}\Omega\Bigr\}}\,.
\end{equation}
In this formula the undetermined normalization constant $\mathrm{\infty}$ was included into the integration measure
$[\mathscr{D}\Sigma]$ and the path integral over $\tilde{\gamma}$'s was converted to the surface functional integral
as was promised:
$$
\int[\mathscr{D}\tilde{\gamma}]\int\limits_{\mathscr{U}_{\tilde{\gamma}}}[\mathscr{D}\Sigma]\cdots\ =
\int\limits_{\bigcup_{\tilde{\gamma}}\mathscr{U}_{\tilde{\gamma}}}[\mathscr{D}\Sigma]\cdots\ =:
\int\limits_\mathscr{U}[\mathscr{D}\Sigma]\cdots\ \,.
$$
The set $\mathscr{U}=\bigcup_{\tilde{\gamma}}\mathscr{U}_{\tilde{\gamma}}$ over which the functional integration is
carried out contains all extended phase space strings which are anchored to the given classical trajectory
$\gamma_{cl}$.

To eliminate Hamiltonian $H$ completely we need to express $\mathbf{S}_{cl}(q_1,t_1\,|\,q_0,t_0)$ in terms of the
force field. Such a quantity may not exist in general, however we will see that in special cases one can recover an
appropriate analog of $\mathbf{S}_{cl}(q_1,t_1\,|\,q_0,t_0)$ requiring a certain behavior of
$$
\mathbf{A}(q_1,t_1\,|\,q_0,t_0)\propto
\mathrm{e}^{\frac{i}{\hslash}\mathbf{S}_{cl}}\int\limits_{\mathscr{U}}[\mathscr{D}\Sigma]\exp{\Bigl\{\frac{i}{\hslash}\int\limits_{\Sigma}\Omega\Bigr\}}\,.
$$
in the limit $\hslash\rightarrow 0$.

\section{Functionality}

A major advantage of the surface functional integral formulation rests in its explicit independence on Hamiltonian $H$.
From the point of view of classical physics, dynamical equations and the force fields entering them seem to be more
fundamental than the Hamiltonian and/or Lagrangian function, which provide these equations in a relatively compact
but ambiguous way, see \cite{guys}. Therefore from the conceptual point of view, Formula (\ref{FeynmanIII}) gives us
transition probability amplitude from a different and hopefully new perspective. It is clear that for the potential generated
forces the surface functional integral formula (\ref{FeynmanIII}) gives nothing new compared to (\ref{FeynmanI}), since
in this case $\Omega$ is closed and can be represented as $\Omega=d(p_adq^a-Hdt)$.
There are, of course, some hidden subtleties which we pass over either quickly or in silence, however, all of them are
discussed in \cite{kochan}.

To show functionality we need to analyze either a strongly non-Lagrangian system \cite{douglas} or a weakly non-Lagrangian
one.
For the sake of simplicity let us focus on the second case. To this end, let us consider a system
consisting of a free particle with unit mass affected by friction:
$$
\ddot{q}=-\kappa\dot{q}\ \ \Leftrightarrow\ \ \Omega=dp\wedge(dq-p\,dt)-\kappa p\,dq\wedge dt\,.
$$
In the considered example, the surface functional integral can be carried out explicitly (for details see
\cite{kochan}). At the end one arrives at the path integral in the configuration space with a surprisingly trivial
result:
$$
\begin{aligned}
\int\limits_{\mathscr{U}}[\mathscr{D}\Sigma]\exp{\Bigl\{\frac{i}{\hslash}\int\limits_{\Sigma}\Omega\Bigr\}}
&\propto\exp{\Bigl\{-\frac{i}{\hslash}\int\limits_{t_0}^{t_1}\bigl(\tfrac{1}{2}\dot{q}_{cl}^2-\kappa p_{cl}q_{cl}\bigr)dt\Bigr\}}\times\\
&\int[\mathscr{D}q]\exp{\Bigl\{\frac{i}{\hslash}\int\limits_{t_0}^{t_1}\bigl(\tfrac{1}{2}\dot{q}^2-\kappa p_{cl}q\bigr)dt\Bigr\}}\,.
\end{aligned}
$$
If we define $\mathbf{S}_{cl}$ to be
\begin{equation}\label{action}
\mathbf{S}_{cl}(q_1,t_1\,|\,q_0,t_0)=\int\limits_{t_0}^{t_1}\bigl(\tfrac{1}{2}\dot{q}_{cl}^2-\kappa p_{cl}q_{cl}\bigr)dt\,,
\end{equation}
then
$$
\mathbf{A}(q_1,t_1\,|\,q_0,t_0)\propto\int[\mathscr{D}q]\exp{\Bigl\{\frac{i}{\hslash}\int\limits_{t_0}^{t_1}\bigl(\tfrac{1}{2}\dot{q}^2-\kappa p_{cl}q\bigr)dt\Bigr\}}
$$
and in the classical limit $\hslash\rightarrow 0$ we arrive at the saddle point equation which is specified by the
functional term in the exponent above:
$$
\ddot{q}=-\kappa\dot{q}_{cl}\,.
$$
This differential equation is different from the equation $\ddot{q}=-\kappa\dot{q}$ that we started with initially, but
both of them coincide when a solution $q(t)$ satisfying $q(t_0)=q_0$ and $q(t_1)=q_1$ is looking for. In the present
situation we gain:
$$
\mathbf{S}_{cl}=\frac{\kappa}{4}(q_1-q_0)\frac{(q_0+3q_1)\mathrm{e}^{-\kappa
t_1}-(q_1+3q_0)\mathrm{e}^{-\kappa t_0}}
{\mathrm{e}^{-\kappa t_0}-\mathrm{e}^{-\kappa t_1}}
$$
and
\begin{equation}\label{propagator}
\mathbf{A}(q_1,t_1|q_0,t_0)=
\sqrt{\frac{\kappa}{4\pi i \hslash\tanh\tfrac{\kappa}{2}(t_1-t_0)}}\,
\mbox{\large{e}}^{\frac{i}{\hslash}\mathbf{S}_{cl}}\,.
\end{equation}
Here we have already employed the normalization conditions specified in the first paragraph. One can immediately verify
that in the frictionless limit $(\kappa\rightarrow 0)$ the transition probability amplitude
$\mathbf{A}(q_1,t_1\,|\,q_0,t_0)$ matches the Schr\"{o}dinger propagator for a single free particle.

\section{Conclusion}

Quantization of dissipative systems has been very attractive problem from the early days of quantum mechanics.
It has been revived again and again across the decades. Many phenomenological techniques and effective
methods have been suggested. References \cite{all others I} and \cite{all others II} provide a very basic list of
papers dealing with this point.

We have developed here a new quantization method that generalizes the conventional path integral approach.
We have focused only on the nonrelativistic quantum mechanics of spinless systems. However, the generalization to
the field theory is straightforward.

Let us stress that the proposed method represents an alternative approach to \cite{all others II} and
possesses several qualitative advantages. For example, propagator (\ref{propagator}) is invariant with respect to time
translations, the same symmetry property which is possessed by the underlying equation of motion. Moreover, it is
reasonable to expect that the ``dissipative quantum evolution'' will not remain Markovian. This fact is again confirmed,
since the probability amplitude under consideration does not satisfy the memoryless Chapman-Kolmogorov equation mentioned
in the first paragraph.\\

Finally, let us believe that the simple geometrical idea behind the surface functional integral quantization will fit
within the Ludwig Faddeev dictum \emph{quantization is not a science, quantization is an art}!

\begin{acknowledgements}
This research was supported in part by M\v SMT Grant LC06002, VEGA Grant 1/1008/09 and  by the M\v S SR program for
CERN and international mobility.\\
\\

\centerline{$\mathscr{A.}$\ \ \ \ \ \ $\mathscr{M.}$\ \ \ \ \ \ $\mathscr{D.}$\ \ \ \ \ \ $\mathscr{G.}$}
\end{acknowledgements}


\begin{thebibliography}{99}
%
\bibitem{feynman1}
R. P. Feynman, A. R. Hibbs: \emph{Quantum Mechanics and Path Integrals}, McGraw-Hill Inc., New York, 1965.
%
\bibitem{feynman2}
R. P. Feynman: \emph{Space-Time Approach to Non-Relativistic Quantum Mechanics}, Rev. of Mod. Phys. \textbf{20} (1948), 367-387.
%
\bibitem{guys}
E. P. Wigner: \emph{Do the Equations of Motion Determine the Quantum Mechanical Commutation Relations?}, Phys. Rev.
\textbf{77} (1950), 711-712.

S. Okubo: \emph{Does the Equation of Motion determine Commutation Relations?}, Phys. Rev. D \textbf{22} (1980), 919-923.

M. Henneaux: \emph{Equations of motion, Commutation Relations and Ambiguities in the Lagrangian Formalism},
Annals Phys. \textbf{140} (1982), 45-64.
%
\bibitem{kochan}
D. Kochan: \emph{Direct quantization of equations of motion}, Acta Polytechnica \textbf{47} No. 2-3 (2007), 60-67, arXiv:hep-th/0703073.

D. Kochan: \emph{How to Quantize Forces(?): An Academic Essay on How the Strings Could Enter Classical Mechanics}, J. Geom. Phys.
60 (2010), 219-229, arXiv:hep-th/0612115.

D. Kochan: \emph{Quantization of Non-Lagrangian systems: some irresponsible speculations} AIP Conf. Proc. \textbf{956} (2007), 3-8.

D. Kochan: \emph{Quantization of Non-Lagrangian Systems}, Int. J. Mod. Phys. A \textbf{24} Nos. 28 $\&$ 29 (2009), 5319-5340.

D. Kochan: \emph{Functional integral for non-Lagrangian systems}, Phys. Rev. A \textbf{81}, 022112 (2010), arXiv:1001.1863.
%
\bibitem{douglas}
J. Douglas: \emph{Solution of the inverse problem in the calculus of variations}, Trans. Am. Math. Soc. \textbf{50} (1941),
71-128.
%
\bibitem{all others I}
H. Beteman: \emph{On Dissipative Systems and Related Variational Principles}, Phys. Rev. \textbf{38} (1938), 815-819.

R. P. Feynman, F. L. Vernon: \emph{The Theory of a General Quantum System Interacting with a Linear Dissipative System},
Annals of Physics \textbf{24} (1963), 118-173.

M. D. Kostin: \emph{On the Schr\"{o}dinger-Langevin Equation}, J. Chem. Phys. \textbf{57} (1972), 3589-3591.

K. K. Kan, J. J. Griffin: \emph{Quantized Friction and the Correspondence Principle: Single Particle with Friction},
Phys. Lett. \textbf{50B} (1974), 241-243.

R. W. Hasse: \emph{On the Quantum Mechanical Treatment of Dissipative Systems}, J. Math. Phys. \textbf{16} (1975),
2005-2011.

H. Dekker: \emph{Classical and quantum mechanics of the damped harmonic oscillator}, Phys. Rep. \textbf{80} (1981), 1-112.

R. Alicki: \emph{Path integrals and stationary phase approximation for quantum dynamical semigroups. Quadratic systems},
J. Math. Phys. \textbf{23} (1982), 1370-1375.

A. O. Caldeira, A. J. Leggett: \emph{Path integral approach to quantum Brownian motion}, Physica \textbf{121A} (1983), 587-616.

J. Geicke: \emph{Semi-classical quantisation of dissipative equations}, J. Phys. A: Math. Gen. \textbf{22} (1989) 1017-1025.

U. Weiss: \emph{Quantum Dissipative Systems} (2$^{\mathrm{nd}}$ ed.), Series in Modern Condensed Matter Physics-Vol.10,
World Scientific, Singapore, 1999.

V. E. Tarasov: \emph{Quantization of non-Hamiltonian and Dissipative Systems}, Phys. Lett. A \textbf{288} (2001), 173-182,
quant-ph/0311159.

H. P. Breuer, F. Petruccione: \emph{The Theory of Open Quantum Systems}, Oxford University Press, Oxford, 2002.

M. Razavy: \emph{Classical and Quantum Dissipative Systems}, Imperial Colleges Press, London, 2005.

D. Chru\'{s}ci\'{n}ski, J. Jurkowski: \emph{Quantum damped oscillator I: dissipation and resonances}, Ann. Phys.
\textbf{321} (2006), 854-874, quant-ph/0506007.

D. M. Gitman, V. G. Kupriyanov: \emph{Canonical quantization of so-called non-Lagrangian systems}, Eur. Phys. J. C \textbf{50} (2007), 691-700,
arXiv: hep-th/0605025.

V. E. Tarasov: \emph{Quantum Mechanics of Non-Hamiltonian and Dissipative Systems}, Elsevier Science, 2008.
%
\bibitem{all others II}
P. Caldirola: \emph{Forze non conservative nella meccanica quantistica}, Nuovo Cim. \textbf{18} (1941), 393-400.

E. Kanai: \emph{On the Quantization of the Dissipative Systems}, Prog. Theor. Phys. \textbf{3} (1948), 440-442.

I. C. Moreira: \emph{Propagators for the Caldirola-Kanai-Schr\" odinger Equation}, Lett. Nuovo Cim. \textbf{23} (1978),
294-298.

A. D. Jannussis, G. N. Brodimas, A. Strectlas: \emph{Propagator with friction in Quantum Mechanics}, Phys. Lett.
\textbf{74A} (1979), 6-10.

U. Das, S. Ghosh, P. Sarkar, B. Talukdar: \emph{Quantization of Dissipative Systems with Friction Linear in Velocity},
Physica Scripta. \textbf{71} (2005), 235-237.
%
\end{thebibliography}
\end{document}